\documentclass[aps,prd,preprintnumbers,nofootinbib,floatfix,floats,groupedaddress,twocolumn]{revtex4}

\usepackage{bm}
\usepackage{latexsym}
\usepackage{dcolumn}
\usepackage{amsmath,amsfonts,amssymb}
\usepackage{graphicx,epsfig}
\usepackage{color}
\usepackage[active]{srcltx}
\usepackage[caption=false]{subfig}
\usepackage{slashed}
\usepackage{tikz}
\usetikzlibrary{shapes,snakes}

\usepackage[mathscr]{euscript}

\def\nn{\nonumber}

\def\l{\left}
\def\r{\right}
\def\DM{\mathrm{d}}

\newcommand{\gae}{\lower 3pt \hbox{$\,\, \buildrel {\scriptstyle >}\over {\scriptstyle
\sim}\,\,$}}
\newcommand{\lae}{\lower 2pt \hbox{$\, \buildrel {\scriptstyle <}\over {\scriptstyle
\sim}\,$}}

\reversemarginpar

\def \lp { \ell_0 }

\begin{document}

\title{BGV theorem, Geodesic deviation, and Quantum fluctuations}

 \author{Dawood Kothawala}
 \email{dawood@iitm.ac.in}
 \affiliation{Department of Physics, Indian Institute of Technology Madras, Chennai 600 036}

\date{\today}
\begin{abstract}
\noindent
I point out a simple expression for the ``Hubble" parameter $\mathscr{H}$, defined by Borde, Guth and Vilenkin (BGV) in their proof of past incompleteness of inflationary spacetimes. I show that the parameter $\mathscr{H}$ which an observer $O$ with four-velocity $\bm v$ will associate with a congruence $\bm u$ is equal to the fractional rate of change of the magnitude $\xi$ of the Jacobi field $\bm \xi$ associated with $\bm u$, measured along the points of intersection of $O$ with $\bm u$, with its direction determined by 
$\bm v$. I then analyse the time dependence of $\mathscr{H}$ and $\xi$ using the geodesic deviation equation, computing these exactly for some simple spacetimes, and perturbatively for spacetimes close to maximally symmetric ones. The perturbative solutions are used to characterise the rms fluctuations in these quantities arising 
due to possible fluctuations in the curvature tensor.
\end{abstract}

\maketitle
\section{Introduction} \label{sec:intro} 
The BGV theorem illustrates past geodesic incompleteness of inflationary universes under very plausible assumptions, without appealing to field equations or energy conditions. Instead, the theorem uses a well motivated kinematical setup to define the ``expansion rate" $\mathscr{H}$ that an observer $O$ will associate with a given congruence in an arbitrary curved spacetime. 

I show here that the BGV expression for $\mathscr{H}$ can be cast completely in terms of the Jacobi fields (deviation vectors) associated with the congruence. This result follows from some elementary manipulations, but as a bonus, it allows us to study the time evolution of expansion using the geodesic deviation equation. 

\begin{figure}[htb!]%
    {{\includegraphics[width=0.3\textwidth]{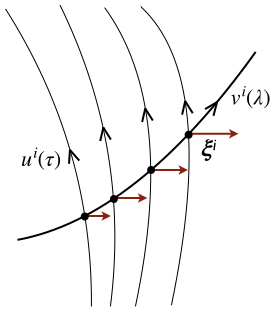} }}%
    \caption{The geometric setup for the BGV theorem. The Jacobi field (the deviation vector) $\bm \xi$ along the intersection points would be parallel to $\bm v_{\rm \perp} = \bm v + (\bm v \bm \cdot \bm u) \bm u = (\gamma v) \bm n$ (see text). See also Fig. \ref{fig:bgv2} in the last section.}%
    \label{fig:bgv}%
\end{figure}

\section{Rewriting $\mathscr{H}$} \label{sec:H}

In \cite{bgv}, the expansion rate of a congruence $\bm u$, as measured by an observer $O$ with four velocity $\bm v$, is defined as
	\begin{eqnarray}
	\mathscr{H} = \frac{\bm n \bm \cdot \l( \bm \nabla_{\bm v} \bm u\ \Delta \lambda \r)}{\bm n \bm \cdot \l( \bm v\ \Delta \lambda\r) } 
	\equiv \frac{\text{projected change in velocity}}{\text{projected distance traversed}}
	\label{eq:bgv-defn}
	\end{eqnarray}
where $\lambda$ is the parameter along $\bm v$, and $\bm n$ is a unit vector orthogonal to $\bm u$, such that 
\begin{equation}
  {\bm v}=\left\{
  \begin{array}{@{}ll@{}}
    \gamma \l( \bm u + v \bm n \r) & \hspace{.25cm} \text{if}\ {\bm v}\ \text{is timelike} \\
    \\
    \gamma \l( \bm u \pm \bm n \r) & \hspace{.25cm} \text{if}\ {\bm v}\ \text{is null} 
  \end{array}\right.
\end{equation} 
with $\gamma=(1-v^2)^{-1/2}$ for the timelike case. (Note that the results for the null case can be obtained from the timelike one by pretending $\gamma$ and $v$ as unrelated and setting $v=\pm1$.)

Starting from the definition Eq.~(\ref{eq:bgv-defn}), and substituting for $\bm v$, 
\begin{equation}
\mathscr{H} = \left\{
  \begin{array}{@{}ll@{}}
    \bm n \bm \cdot \bm \nabla_{\bm n} \bm u + \frac{1}{v} \bm n \bm \cdot \bm a_{(\bm u)} & \hspace{0.25cm} \text{if}\ {\bm v}\ \text{is timelike} \\
    \\
    \bm n \bm \cdot \bm \nabla_{\bm n} \bm u \pm \bm n \bm \cdot \bm a_{(\bm u)} & \hspace{0.25cm} \text{if}\ {\bm v}\ \text{is null} 
  \end{array}\right.
\end{equation} 
where $\bm a_{(\bm u)} = \nabla_{\bm u} \bm u$ is the acceleration of $\bm u$. At this point, we note that $\mathscr{H}$ is being measured by the observer $O$ at each of its intersections with $\bm u$. \textit{Therefore, there is a one to one correspondence between $\bm n$, and the deviation vector $\bm \xi$ that connects two nearby geodesics of the congruence which $O$ intersects in succession}. Write this deviation vector as $\bm \xi = \xi \bm n$. We now restrict to geodesic congruences ($\bm a_{(\bm u)}=0$), and use the standard condition on the deviation vector: $\mathscr{L}_{\bm u} \bm \xi=0$
\footnote
{For an accelerated congruence, one can not satisfy both $\bm \xi \bm \cdot \bm u=0$ and $\mathscr{L}_{\bm u} \bm \xi=0$ simultaneously, due to the identity 
$\bm \xi \bm \cdot \bm a = \nabla_{\bm u} \l(\bm \xi \bm \cdot \bm u\r) - \bm u \bm \cdot \mathscr{L}_{\bm u} \bm \xi$. Nevertheless, a suitable choice of $\mathscr{L}_{\bm u} \bm \xi$ can be made for an accelerated congruence so as to give the same result. We will, however, not discuss this case further.
}
. This implies $\mathscr{L}_{\bm u} \bm n = - \bm n \nabla_{\bm u} \ln \xi$, and therefore
	\begin{eqnarray}
	\nabla_{\bm n} \bm u = \nabla_{\bm u} \bm n + \bm n \nabla_{\bm u} \ln \xi
	\end{eqnarray}
which immediately gives 
\begin{equation}
\mathscr{H} =  \nabla_{\bm u} \ln \xi \hspace{0.5cm} (\text{for}\ {\bm v}\ \text{timelike or null})
  	\label{eq:H}
\end{equation} 
The above relation then gives the expansion $\mathscr{H}$ as defined by BGV in terms of first derivative of deviation vectors at the intersection points.

\underline{\textit{The BGV expression}}:

If one instead starts from the definition Eq.~(\ref{eq:bgv-defn}) and substitute for $\bm n$, we obtain
	\begin{equation}
	\mathscr{H} = \left\{
	  \begin{array}{@{}ll@{}}
	    \nabla_{\bm v} \l( \ln \sqrt{\frac{\gamma+1}{\gamma-1}} \r) - \frac{1}{\gamma^2-1} \bm u \bm \cdot \bm a_{(\bm v)} &\hspace{0.25cm}  \text{if}\ {\bm v}\ \text{is timelike} \\
	    \\
	   \nabla_{\bm v} \frac{1}{\gamma} \, - \frac{1}{\gamma^2}\ \bm u \bm \cdot \bm a_{(\bm v)} & \hspace{0.25cm} \text{if}\ {\bm v}\ \text{is null} 
	  \end{array}\right.
	\end{equation} 
%
	where $\bm a_{(\bm v)} = \nabla_{\bm v} \bm v$ is the acceleration of $\bm v$. For $\bm a_{(\bm v)} = 0$, the above expression reduces to the one derived in BGV. 
	It is worth noting that the expression for $\mathscr{H}$ can be expressed either as a total derivative along $\bm u$, or a total derivative along $\bm v$ (for geodesic $
	\bm v$). The analysis of BGV proceeds by defining 
	$$\mathscr{H}_{\rm avg} = \frac{1}{\Delta \lambda} \int \limits_{\lambda_i}^{\lambda_f} \mathscr{H} \DM \lambda$$
	($\Delta \lambda = \lambda_f-\lambda_i$) and proving that, if $\mathscr{H}_{\rm avg} > 0$ for a congruence that is almost comoving with a geodesic observer at late 
	times ($\gamma_f \approx 1$), then $\Delta \lambda$ is bounded above implying that $\bm v$ can not be extended indefinitely to the past, and hence is incomplete.

Given the alternate expression for $\mathscr{H}$ derived here, we may now run an argument similar to the one above. We describe this next.

\subsubsection{Geodesic incompleteness of the congruence} \label{sec:conseq}

Following BGV, consider the average
	\begin{eqnarray}
	\mathscr{H}_{\rm avg} = \frac{1}{\Delta \tau} \int_{\tau_i}^{\tau_f} \mathscr{H} \DM \tau
	\end{eqnarray}
where $\tau$ is the proper time along $\bm u$. This now results in
	\begin{eqnarray}
	\frac{\xi_f}{\xi_i} = \exp \l[ \mathscr{H}_{\rm avg} \Delta \tau \r]
	\end{eqnarray}
Note that the above expression gives an intuitively natural interpretation of the condition $\mathscr{H}_{\rm avg}>0$ imposed by BGV; under these conditions, the observer will measure greater deviations at later intersections with the congruence, since $\xi_f > \xi_i$. 

Now suppose that $\xi$ is bounded above by, say $1/\sqrt{\Lambda}$, and also below by, say, $\lp$. 
\\
 \line(1,0){250}
\\
\textbf{{Comment\,$\bm 1$}:} Though the upper bound might come from size of the observable universe, no direct connection is possible without a more precise characterisation of the causal structure. The lower bound is expected to come from quantum gravitational fluctuations, and their effects on (de-)focussing of geodesics \cite{ml1, ml2, ford-etal1, ford-etal2, ford-etal3}.
\\
 \line(1,0){250}

Assuming such bounds exist, $\xi_f < 1/\sqrt{\Lambda}$ and $\xi_i > \lp$, so that 
	\begin{eqnarray}
	\frac{\xi_f}{\xi_i} < \frac{1}{\sqrt{\Lambda \lp^2}}
	\end{eqnarray}
This immediately implies, for $\mathscr{H}_{\rm avg}>0$,
	\begin{eqnarray}
	\Delta \tau < \frac{1}{2 \mathscr{H}_{\rm avg} } \ln \l( \frac{1}{\Lambda \lp^2} \r)
	\end{eqnarray}
The implications of the above result are:
	\begin{enumerate}
	\item For $\Lambda \lp^2 = 0$, the congruence would be geodesically complete. This would happen if either $\Lambda=0$ or $\lp=0$.
	\item For $\Lambda \lp^2<1$, the congruence would be geodesically incomplete and can not be extended beyond a proper time $$\Delta \tau_{\rm max}=- (2 \mathscr{H}_{\rm avg})^{-1} \ln \l( \Lambda \lp^2 \r)$$
	\end{enumerate}	

 \line(1,0){250}
\\
\textbf{{Comment\,{$\bm 2$}}:} I must emphasise that the above points pertain only to geodesic (in)completeness of the congruence $\bm u$. The original BGV theorem still applies, of course, for the geodesic observer $\bm v$, implying incompleteness. 
\\
 \line(1,0){250}

Finally, since the expression for $\mathscr{H}$, Eq.~(\ref{eq:H}), depends on $\bm u$, one might ask what role the observer plays in the analysis, and one may ask
\subsubsection{What is the role of the observer?} \label{sec:observer-role}
This is easily clarified by noticing that, the vector field 
$\bm n$ is determined by the observer four-velocity $\bm v$; specifically, $\bm n = \l( \gamma v \r)^{-1} \bm v - v^{-1}\bm u$. Therefore, at any given intersection point of 
$\bm v$ with $\bm u$, the direction of the Jacobi field $\bm \xi$ is determined by the observer's motion. That is, the observer's motion determines which member of the congruence it is going to intersect next, and $\bm \xi$ is then the deviation vector between these two members of the congruence. Now, since the time evolution of $\xi$ (Eq.~(\ref{eq:xi-evolution}) below) depends on $\bm n$, the corresponding solution will depend on the observer. (While this is clear from the equations, it will be instructive to have an explicit example that demonstrates the point.) Such a dependence is also evident from the curvature term $R_{abcd} u^a n^b u^c n^d$ that appears in the evolution equations (next section), since this term can alternately be written as $(\gamma v)^{-2} R_{abcd} u^a v^b u^c v^d$. Fig. (\ref{fig:bgv2}) further elaborates on the role of the observer.

The role of the observer can also be clarified by considering the following example in Minkowski spacetime. The straight lines in Minkowski coordinates define an inertial congruence, while curves of constant acceleration $a$ define the Rindler congruence. One may then choose $\bm u$ to be given by one of these congruence, with a specific member of the other congruence as the observer with four velocity $\bm v$. It is trivial to compute $\mathscr{H}$ for the above two cases from the given expressions (with non-zero accelerations), and the result turns out to be:

(a) Inertial congruence, Rindler observer: $$\mathscr{H}=0$$ 

(b) Rindler congruence, Inertial observer: $$\mathscr{H}(T)=-\frac{a(T)}{v_{\rm rel}(T)}$$

In (b), the quantities $a(T)$ and $v_{\rm rel}(T)$ are, respectively, the acceleration of the member of Rindler congruence at each point of intersection $T$ with the observer, 
and $v_{\rm rel}(T)$ is the relative velocity computed from $\gamma_{\rm rel}=-\bm u \bm \cdot \bm v$. The above example is simple and illuminating in highlighting the role of the observer. As a curious aside, note that, (b) can be re-written as $|a| = |\mathscr{H}| \times |v_{\rm rel}|$, which is in stark contrast with the usual definition of 
$\mathscr{H}$ in cosmology, given by ``velocity = $H \times$ distance".

\section{Evolution of $\mathscr{H}$ and $\xi$} \label{sec:Hdot}
Having expressed $\mathscr{H}$ in terms of a deviation vector, one can use the geodesic deviation equation (GDE)
	\begin{eqnarray}
	\bm \nabla_{\bm u} \bm \nabla_{\bm u} \xi^i = R^i_{\phantom{i}abm} u^a u^b \xi^m 
	\end{eqnarray}
	to evaluate the derivative of $\mathscr{H}$ along $\bm u$, which is expected to be of interest as a measure of acceleration associated with the expansion of the congruence by an arbitrary observer. We will be mostly interested in the equation governing the magnitude of $\bm \xi$, defined by $\bm \xi = \xi \bm n$. A straightforward computation (see Appendix \ref{app:dev-eq}) gives
	\begin{eqnarray}
	\frac{\bm \nabla^2_{\bm u} \xi}{\xi} = - R_{abcd} u^a n^b u^c n^d + \l( \bm \nabla_{\bm u} \bm n \r)^2 
	\label{eq:xi-evolution}
	\end{eqnarray}	
	which leads to the following equation for $\mathscr{H}$,
        \begin{equation}
        		\bm \nabla_{\bm u} \mathscr{H} = - R_{abcd} u^a n^b u^c n^d + \l( \bm \nabla_{\bm u} \bm n \r)^2 - \mathscr{H}^2
                \label{eq:hubble-evolution}
	\end{equation}
which is a Riccati type differential equation. It is also worth quoting that, when $\bm \nabla_{\bm u} \bm n=0$, one can write
	\begin{equation}
	\epsilon_{\mathscr{H}} = - \frac{\dot{\mathscr H}}{\; \; {\mathscr H}^2} = 1 + {\mathscr H}^{-2} R_{abcd} u^a n^b u^c n^d
         \label{eq:slow-roll}
	\end{equation}
The quantity $\epsilon_{\mathscr{H}}$ is, of course, motivated by the (Hubble) slow roll parameter used in inflationary cosmology. 	

 \line(1,0){250}
\\
\textbf{{Comment\,{$\bm 3$}}:} \textit{Comparison with Raychaudhuri equation}:
 The equation for geodesic deviation is very closely connected with the Raychaudhuri equation which determines the evolution of the expansion scalar $\Theta=\bm \nabla \bm \cdot \bm u$.
	 Defining $H_{\Theta} = \Theta/(D-1)$ where $D$ is the dimension of spacetime, the Raychaudhuri equation becomes
	\begin{equation}
	\bm \nabla_{\bm u} \mathscr{H}_{\Theta} = - \frac{1}{(D-1)} R_{ab} u^a u^b - \frac{1}{(D-1)}\l( \sigma^2 - \omega^2 \r) - \mathscr{H}_\Theta^2
	\label{eq:hubble-rayc}
	\end{equation}
	One can then compare Eq.~(\ref{eq:hubble-rayc}) and Eq.~(\ref{eq:hubble-evolution}), after ignoring shear and rotation in the former and $\l( \bm \nabla_{\bm u} \bm n \r)^2$ in the latter. In 
	this case, Eq.~(\ref{eq:hubble-evolution}) reduces to Eq.~(\ref{eq:hubble-rayc}) if one averages over the directions $\bm n$ that is determined by the observer; that is, 
	upon making the replacement $n^a n^b \to h^{ab}/(D-1)$, with $h_{ab} = g_{ab} + u_a u_b$.
\\
 \line(1,0){250}

We will now analyse some features of the above differential equation, and it's solution, for both $\mathscr{H}$ and $\xi$. Before proceeding to discuss perturbative solutions in general case, we will discuss some simple exact solutions for important special cases. The examples considered (FLRW, maximally symmetric spacetimes, Schwarzschild) should shed light on the similarities and differences between characterising expansion of a congruence {\it a la} BGV versus the usual one that is based on the Raychaudhuri equation. In particular, the cases of perturbed FLRW universe and Schwarzschild spacetime bring into focus the fundamental point regarding what it means to say that a ``spacetime" or a ``space" is expanding (or contracting).
 
\subsection{Exact solutions; special spacetimes} \label{sec:exact-sols}

\underline{\textit{1. Maximally symmetric spacetimes}:} Next, let us consider maximally symmetric spacetimes characterised by 
	\begin{eqnarray}
	R_{abcd} &=& \frac{R}{D (D-1)} \l( g_{ac} g_{bd} - g_{ad} g_{bc} \r)
	\nn \\
	&:=& \Lambda \l( g_{ac} g_{bd} - g_{ad} g_{bc} \r) 
	\end{eqnarray}
where $R=D(D-1) \Lambda$ is the Ricci scalar, $\Lambda$ being constant. It then follows that $R_{abcd} u^a n^b u^c n^d = - \Lambda$, and Eq.~(\ref{eq:xi-evolution}) becomes (with $\ddot \xi=\DM^2 \xi/\DM \tau^2$)
	\begin{eqnarray}
	\ddot \xi =  \Lambda \xi
	\end{eqnarray}	
whose solutions are linear combinations of $e^{\pm \sqrt{\Lambda} \tau}$ for $\Lambda>0$, and of $e^{\pm i \sqrt{\Lambda} \tau}$ for $\Lambda<0$. The ``Hubble" parameter ${\mathscr H}$ is then determined from these easily. 

\textbf{$\bm{\Lambda>0}$:} It is easy to use show that the differential equation Eq.~(\ref{eq:hubble-evolution}) for $\mathscr{H}$ has the solution
	\begin{eqnarray}
	\mathscr{H}(\tau) &=&  \sqrt{\Lambda} \tanh [q(\tau)]
	\nn \\
	q(x) &=& \sqrt{\Lambda} x + \tanh^{-1}\l( \frac{\mathscr{H}(0)}{\sqrt{\Lambda}} \r)
	\label{eq:H-max-symm}
	\end{eqnarray}
with the limits
	\begin{eqnarray}
	\lim \limits_{\tau \to \infty} \mathscr{H}(\tau) &=& \sqrt{\Lambda}
	\nn \\
	\lim \limits_{\tau \to 0} \mathscr{H}(\tau) &=&  \mathscr{H}(0)
	\end{eqnarray}
The corresponding solution for $\xi(\tau)$ is then easily found from 
	\begin{eqnarray}
	\xi(\tau) &=& \xi(0) \exp \l( \int_0^\tau \mathscr{H}(y) \DM y \r)
	\nn \\
	&=& \xi(0) \l[ \cosh\l(\sqrt{\Lambda} \tau\r) + \frac{\mathscr{H}(0)}{\sqrt{\Lambda}} \sinh\l(\sqrt{\Lambda} \tau \r) \r]
	\end{eqnarray}

\textbf{$\bm{\Lambda=0}$:} This is the case of Minkowski spacetime, and the results are easily obtained by taking $\Lambda \to 0$ limit of the previous ones, to give the expected expressions
	\begin{eqnarray}
	\mathscr{H}(\tau) &=& \mathscr{H}(0) 
	\nn \\
	\xi(\tau) &=& \xi(0) + \dot{\xi}(0) \tau
	\end{eqnarray}
	
Finally, let us also point out the following result which follows from Eq.~(\ref{eq:slow-roll}): 
	\begin{equation}
	\epsilon_{\mathscr{H}} = - \frac{\dot {\mathscr H}}{\; \; {\mathscr H}^2} = 1 - \Lambda {\mathscr H}^{-2}
	\end{equation}
	and hence $\lim \limits_{\tau \to \infty} {\epsilon_{\mathscr{H}}} = 0$.

\underline{\textit{2. Perturbed FLRW models}:}

For the unperturbed FLRW spacetime described by the metric $\DM s^2 = - \DM t^2 + a(t)^2 \DM \ell^2$ (where $\DM \ell^2$ is the spatial metric), we have $\xi \propto a(t)$ it is obvious that $\mathscr{H} = \dot a/a$, which coincides with the standard definition of the Hubble parameter. To understand the difference from the standard definition, let us consider a model of {\it perturbed} FLRW spacetime. For simplicity, we will work in two dimensions and choose the synchronous gauge: $\DM s^2 = - \DM t^2 + a(t)^2 \l( 1 + h(t, x) \r) \DM x^2$. In this case, to first order, $\mathscr{H} = {\dot a}/{a} + ({1}/{2}) \dot h$, which is as expected.

\underline{\textit{3. Schwarzschild spacetime}:}

Schwarzschild geometry provides the most unexpected case for which the notion of a ``Hubble" expansion can been considered. But we can nevertheless apply the mathematical set-up to this case. For simplicity, I will focus on radial geodesics, starting from rest from $r=R$. The relevant expressions for the trajectory can be found, for example, in \cite{chandra-mtbh}. A detailed but straightforward algebra (see Appendix \ref{app:schw}) then yields the following differential equation for $\xi(\eta)$
	\begin{equation}
	\xi'' + \tan (\eta/2) \; \xi' - \sec^2(\eta/2) \; \xi = 0
	\label{eq:xi-schw}
	\end{equation}
where $\eta = 2 \arccos \sqrt{r/R}$. The proper time along the geodesics is related to $\eta$ by $\tau/a = (1/2)(R/a)^{3/2} (\eta + \sin \eta)$, where $a=2GM/c^2$ is the Schwarzschild horizon radius. The above equation has an exact solution, which can be written as
	\begin{equation}
	\xi(\eta) = \frac{1}{4} \xi(0) \Biggl[
	5 -  \cos \eta +
	 \left(\frac{8 \dot \xi(0) R^{3/2}}{\xi(0) \sqrt{a}} + 3 \eta \right) 	\tan (\eta/2) \Biggl]
	\end{equation}
	where $\dot \xi(0)=(\DM \xi/\DM \tau)_{\tau=0}$. It then follows that 
	\begin{eqnarray}
	\mathscr{H} &=& \frac{\DM \ln \xi}{\DM \tau}
	\nn \\
	&=& 
        \frac{\sqrt{a}}{2 R^{3/2}} 
        \frac{\xi(0)}{\xi(\eta)}
	\frac{\frac{8 \dot \xi(0) R^{3/2}}{\xi(0) \sqrt{a}} + 3 \eta + (4+\cos \eta)\sin \eta}{(1+\cos \eta)^2}
	\end{eqnarray}
	Two interesting limits of this are
	\begin{eqnarray}
	\lim \limits_{r \to R} \mathscr{H} &=& \frac{\dot \xi(0)}{\xi(0)}
	\nn \\
	\lim \limits_{r \to 0} \mathscr{H} &=& \infty
	\end{eqnarray}

\begin{figure}[htb!]%
    \centering
    {{\includegraphics[width=0.45\textwidth]{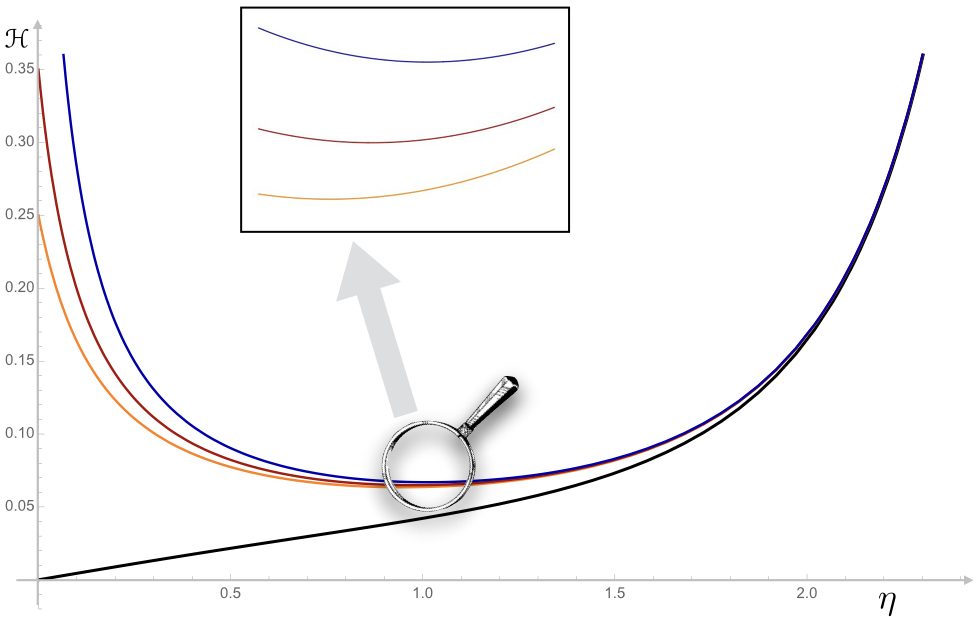} }}%
    \caption{$\mathscr{H}$ as a function of $\eta$, for different values of 
    {\small $\dot \xi(0)=0$ (black), $0.25$ (orange), $0.35$ (red), $0.75$ (blue).}
    }%
    \label{fig:Hschw}%
\end{figure}

The plot in Fig. \ref{fig:Hschw} shows the behaviour of $\mathscr{H}$: it decreases initially till some radius, and then increases, except when $\dot \xi(0)=0$. It is also interesting to consider the case when $\dot \xi(0)<0$. That is, the initial (relative) separation between the geodesics is decreasing. In this case, one obtains, from the relevant plots and a little bit of numerical analysis, that $\xi \to 0$ whenever $|\dot \xi(0)| \gtrsim 1.2 \xi(0) \sqrt{a}/R^{3/2}$; these zeroes of $\xi$ all occur for $\eta \lesssim 2.15$. If $|\dot \xi(0)| \lesssim 1.2 \xi(0) \sqrt{a}/R^{3/2}$, the geodesics would hit the singularity ($r=0$) at $\eta=\pi$ before they cross. 

\subsection{Perturbative solution; generic spacetimes} \label{sec:pert-sols}

\begin{figure}[htb!]%
    {{\includegraphics[width=0.45\textwidth]{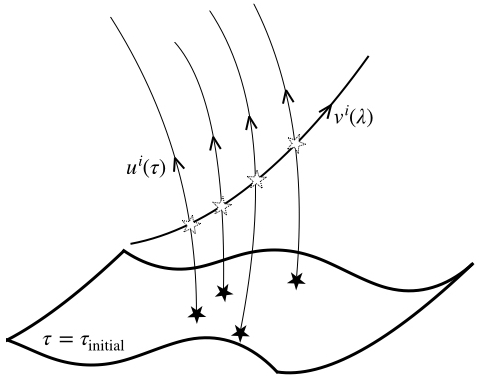} }}%
    \caption
    {
    At each of its intersections $\lambda_\star$ with the congruence, the observer $\bm v$ determines a unit vector $\bm n_\star$ through 
    $\bm v_{\rm \perp} = (\gamma v) \bm n_\star$ (see Fig.\ \ref{fig:bgv} and the discussion around it.) One may then parallel transport $\bm n_\star$ along the 
    corresponding member of the congruence, thereby determining 
    $\bm n$ everywhere at ${\tau_{\rm initial})}$. One then has $\nabla_{\bm u} \bm n=0$ by construction.
    }%
    \label{fig:bgv2}%
\end{figure}

We will now discuss perturbative solutions for general congruences in general spacetimes by making the following assumptions:
\begin{enumerate}
\item The Riemann tensor has the form
	\begin{equation}
	R_{abcd} = \Lambda \l( g_{ac} g_{bd} - g_{ad} g_{bc} \r) + {\sf r}_{abcd}
	\end{equation}
	with $\Lambda \geq 0$ and $||{\sf r}_{abcd}|| \ll \Lambda$ under some suitable norm. 
\item The term $\l( \bm \nabla_{\bm u} \bm n \r)^2$ either vanishes or can be ignored
\footnote
{Note that, since $\bm u \bm \cdot \bm \nabla_{\bm u} \bm n = 0 = \bm n \bm \cdot \bm \nabla_{\bm u} \bm n$, $\bm \nabla_{\bm u} \bm n$ solely lives in the $(D-2)$ space with normals $\bm u$ and $\bm n$. When no special direction exists in this subspace, this term will be zero. See also the discussion in Fig. (\ref{fig:bgv2}).
}. 
\end{enumerate}
Under the above assumptions, the differential equations governing $\mathscr{H}$ and $\xi$, Eqs.~\ref{eq:H}, \ref{eq:hubble-evolution}, become:
\begin{eqnarray}
\mathscr{\dot H} + \mathscr{H}^2 - \Lambda + {\sf r}(\tau) &=& 0
\end{eqnarray}
where $\mathscr{\dot H} = \DM \mathscr{H}/\DM \tau$, ${\sf r}(\tau)={\sf r}_{abcd} u^a n^b u^c n^d$. From the solution of the above equation, one then obtains $\xi(\tau)$ as
\begin{eqnarray}
\xi(\tau) = \xi(0) \exp{\l[\int \limits_0^\tau \DM \tau' \, \mathscr{H}(\tau') \r]}
\end{eqnarray}
It is straightforward to show that the solution is given, to $O([{\sf r}_{abcd}]^2)$, by
\begin{eqnarray}
\mathscr{H}(\tau) &=& \sqrt{\Lambda} \tanh q(\tau) - \int \limits_0^\tau \DM \tau' \, {\sf r}(\tau') \mathcal{F}(\tau', \tau)
\nonumber \\
\xi(\tau) &=& \xi(0) \frac{\cosh q(\tau)}{\cosh q(0)} \Biggl[ 1 - \int \limits_0^\tau \DM \tau'  \int \limits_0^{\tau'} \DM \tau'' \, {\sf r}(\tau'') \mathcal{F}(\tau'', \tau') \Biggl]
\nonumber \\
\mathcal{F}(\tau, \tau') &=& \l( \frac{\cosh q(\tau)}{\cosh q(\tau')} \r)^2
\label{eq:pert-soln-full}
\end{eqnarray}
where $q(x)$ is defined in Eq.~(\ref{eq:H-max-symm}).

The above solutions then allow us to compute the BGV expansion of a congruence as measured by an observer.

\section{Quantum fluctuations} \label{sec:quantum-fluc}
The original analysis by BGV was primarily concerned with the question of geodesic incompleteness on a purely kinematical level, without referring to any sort of energy conditions. The spacetime is assumed to remain classical throughout. The theorem therefore has nothing to say about how quantum effects might affect the conclusion. It is, of course, interesting to probe deeper the fate of geodesic incompleteness, as described by BGV theorem, in a complete theory that incorporates quantum fluctuations of matter fields as well as spacetime. In Sec.~\ref{sec:conseq}, while discussing the issue of geodesic incompleteness of the congruence $\bm u$, the effects of quantum fluctuations were incorporated through the (covariantly defined) lower bound $\lp$ on distances. The existence of a lower bound on geodesic intervals seems to a be a generic consequence of combining principles of GR and quantum mechanics, and expected to be independent of any specific model/framework of quantum gravity. A mathematical formalism that incorporates this into the small structure of spacetime is presented/developed in some recent work \cite{ml1, ml2}. However, as has already been alluded to in \textbf{Comment\,1}, any precise relation along these lines must necessarily involve also the discussion on how such bounds affect the causal structure of spacetime. In absence of this, the discussion can not be carried any further.

There is, however, an alternate, interesting, way to address the issue of quantum fluctuations by treating Eq.~(\ref{eq:hubble-evolution}) as a Langevin equation sourced by a fluctuating Riemann tensor. The fluctuations might arise due to matter fields via the fields equations, or due to quantum nature of spacetime itself. Such an analysis can be done along the lines of \cite{ford-etal1, ford-etal2, ford-etal3}, and we will have more to comment on these works below. At least in linearised gravity, this would yield a specific scaling of quantum fluctuations $\Delta \mathscr{H}$ in $\mathscr{H}$ with $\lp$. Needless to say, such a scaling acquires importance not only from an observational point of view, but also in understanding better the so called cosmological constant problem. Indeed, as is evident from the discussion in Sec.~\ref{sec:conseq}, the parameter $\Lambda \lp^2$ plays an important role in the issue of geodesic (in)completeness (of the congruence).

We will restrict ourselves to linear order in fluctuations in curvature about a maximally symmetric spacetime with $\Lambda>0$. In other words, we will consider $r_{abcd}$ as a stochastic variable in the solutions derived in Section \ref{sec:pert-sols}. Note that, since only $r=r_{abcd} u^a n^b u^c n^d$ appears in the solutions, we may as well treat $r$ as the stochastic variable. Using Eqs.~(\ref{eq:pert-soln-full}), it is then straightforward to show that
	\begin{widetext}
	\begin{eqnarray}
	\left \langle \l( \Delta \mathscr{H} \r)^2 \right \rangle &=& 
	\int \limits_0^\tau \DM \tau' \int \limits_0^\tau \DM \tau'' 
	\mathcal{C}[\tau', \tau'']
	\mathcal{F}(\tau', \tau) \mathcal{F}(\tau'', \tau)
	\nn \\
	\nn \\
	\left \langle \l( \Delta \mathscr{\xi} \r)^2 \right \rangle &=& \xi(0)^2 \mathcal{F}(\tau, 0) 
	\int \limits_0^\tau \DM \tau_1'  \int \limits_0^{\tau_1'} \DM \tau_1'' \int \limits_0^\tau \DM \tau_2' \int \limits_0^{\tau_2'} \DM \tau_2'' \,
	\mathcal{C}[\tau_1'', \tau_2'']
	\mathcal{F}(\tau_1'', \tau_1') \mathcal{F}(\tau_2'', \tau_2')
	\label{eq:rms-flucs}
	\end{eqnarray}
	\end{widetext}
where $\mathcal{C}[\tau, \tau']=\l \langle \Delta {\sf r}(\tau) \Delta {\sf r}(\tau') \r \rangle$, and $\Delta \mathcal{O}=\mathcal{O} - \langle \mathcal{O} \rangle$. 

Any further computation requires knowledge of the correlator $\mathcal{C}[\tau, \tau']$. However, some generic scaling limits can be obtained for the rms fluctuations defined by $\Delta \mathcal{O}_{\rm rms} = {\l \langle \l(\Delta O\r)^2 \r \rangle}^{1/2}$.

\textbf{1. Short time behaviour; $\tau \to 0$}:
	\begin{eqnarray}
	\Delta \mathscr{H}_{\rm rms} &=& \sqrt{\mathcal{C}[0,0]} \, \tau + O(\tau^2)
	\nn \\
	\Delta \xi_{\rm rms} &=& \frac{1}{2} \sqrt{\mathcal{C}[0,0]} \, \xi(0) \tau^2 + O(\tau^3)
	\label{eq:rms-flucs-short-time}
	\end{eqnarray}	
	assuming $\mathcal{C}[0,0]$ is finite.
	
\textbf{2. Long time behaviour; $\tau \to \infty$}:
	\begin{eqnarray}
		\Delta \mathscr{H}_{\rm rms} &\propto& \left\{
	  \begin{array}{@{}ll@{}}
	    e^{-2 \sqrt{\Lambda} \tau} & \hspace{.25cm} \text{if}\ \Lambda>0 \\
	    \\
	    \tau^{-2} & \hspace{.25cm} \text{if}\ \Lambda=0
	  \end{array}\right.
	\nn \\
	\nn \\
	\Delta \xi_{\rm rms} &\propto& \left\{
	  \begin{array}{@{}ll@{}}
	    e^{\sqrt{\Lambda} \tau} & \hspace{.615cm} \text{if}\ \Lambda>0 \\
	    \\
	    \tau & \hspace{.615cm} \text{if}\ \Lambda=0
	  \end{array}\right.
	  \nn \\
	\label{eq:rms-flucs-long-time}
	\end{eqnarray}	
	assuming all the relevant integrals are bounded in this limit. 
	
Note that the results presented above, specifically the time scaling limits, are very general and should apply to any model of curvature fluctuations as long as the stated assumptions are satisfied. Several specific models for such fluctuations have been considered in the literature, and the most recent application of these is \cite{ford-etal1}, which presents an analysis similar to the one in this section. I comment on this paper in the Appendix \ref{app:comment}.

\section{Concluding remarks} \label{sec:remarks}
In their insightful paper, BGV had provided a kinematical condition under which a geodesic worldline, if it measures an average positive expansion of an almost comoving congruence, can be shown to be past incomplete. In this paper, we have derived an alternate expression for the BGV expansion $\mathscr{H}$ in terms of Jacobi fields associated with the congruence. Besides the implications discussed in the main text, a proper definition of $\mathscr{H}$ is relevant to address the crucial question: 
{What does it mean to say that a ``spacetime" is expanding}? As was emphasised already in \cite{bgv}, it is more fruitful to talk of expansion as a property of a congruence measured by some observer. Our example based on radial geodesic congruence in $1+1$ Schwarzschild illustrates this point particularly nicely. Cast in terms of deviation vectors, our analysis contrasts the usual discussion of expansion through Raychaudhuri equation with the one based on the physically more appealing construction 
of BGV. 

To obtain the perturbative solution for $\mathscr{H}$, we made the assumption that $\nabla_{\bm u} \bm n$ term in the differential equation can be ignored. While this assumption does enable us to get analytical solutions that yield considerable insights, it would be worth having a characterisation of how it contributes to the final results. However, to do so would require specifying a specific observer trajectory, and must presumably be done on a case by case basis numerically.

Having obtained a perturbative solution for $\mathscr{H}$ and $\xi$, we considered the issue of quantum fluctuations, essentially along the lines of Refs. \cite{ford-etal2, ford-etal3, ford-etal1}, by treating the curvature term $r$ as a stochastic variable. Note that, in terms of the congruence and observer four-velocities, 
${\sf r}-\Lambda= (\gamma v)^{-2} R_{abcd} u^a v^b u^c v^d$, and is essentially the sectional curvature of the plane $\bm u \bm \wedge \bm v$:
$$
K[\bm u \bm \wedge \bm v] = \frac{R_{abcd} u^a v^b u^c v^d}{{\bm u}^2 {\bm v}^2 - (\bm u \bm \cdot \bm v)^2} = {\sf r}-\Lambda
$$
We may therefore say that the rms fluctuations are essentially sourced by fluctuations in $K[\bm u \bm \wedge \bm v]$. A proper physical interpretation of this could be illuminating.

We must also point out a conceptual subtlety concerning the treatment of fluctuations. We have computed fluctuations involving terms like 
$\left \langle \mathscr{H}(\tau) \mathscr{H}(\tau) \right \rangle$. Strictly speaking, what is relevant from the point of view of the observer $O$ are the correlations at two subsequent points of intersections of $O$ with the congruence. Technically, this is easy to take care of in the above expressions by simply distinguishing the $\bm u$'s appearing in terms like $\left \langle {\sf r}(\tau) {\sf r}(\tau') \right \rangle$, and using appropriate (different) limits of integrations for $\tau$ and $\tau'$. However, nothing more general can be done beyond this unless a congruence and the observer are explicitly specified.

Finally, some of our results concerning fluctuations might be relevant for the cosmological constant problem, and it is worth probing them further along these lines; see, for example, \cite{paddy-cc-flucs} for a discussion on the role of fluctuations. For some recent work that discuss the cosmological constant problem along these lines, the one most immediately relevant for the discussion here is the work by Wang, Zhu, and Unruh \cite{unruh-wang-cc-flucs}. This recent work is different from previous ones in that it considers both the mean value and the fluctuations of the energy density, unlike, for instance, \cite{paddy-cc-flucs} which only deals with the fluctuation part. The effective macroscopic $H$ nevertheless turns out to be small. (A related attempt is \cite{carlip-cc-flucs}, which also differs from \cite{unruh-wang-cc-flucs}; 
see \cite{unruh-wang-carlip-comment}).

Since we have managed to relate the BGV expansion $\mathscr{H}$ to the fractional rate of change of the deviation vector as measured by an arbitrary observer, our analysis provides a more general set-up to study expansion and fluctuations in arbitrary curved spacetimes, not restricted to FLRW models. A similar analysis has been presented in \cite{unruh-wang-cc-flucs2}, and it will be worth checking how the results of \cite{unruh-wang-cc-flucs, unruh-wang-cc-flucs2} work in our formalism. Despite some essential technical differences, one might investigate the $\Lambda \to \infty$ of $\Delta \mathscr{H}_{\rm rms}$ of our results, since the key claim of aforementioned papers is that large vacuum energy predicted by QFT computations can lead to a small effective cosmological constant. This seems to require more work since one needs to know details of the correlator $\mathcal{C}[\tau, \tau']$ for arbitrary $\tau, \tau'$. However, if the coincidence limit of $\mathcal{C}[\tau, \tau']$ is rendered finite by the same cut-off (corresponding to length $1/\sqrt{\Lambda}$) that makes vacuum energy divergent \cite{ml1, ml2}, then some very generic considerations can give the leading behaviour of $\left \langle \l( \Delta \mathscr{H} \r)^2 \right \rangle$. A model calculation along these lines will be presented elsewhere, connecting also with the results in \cite{unruh-wang-cc-flucs}.

Finally, given some information about the curvature correlators - for instance, coming from some quantum gravitational considerations - would facilitate the discussion of quantum structure of spacetime and singularities. 

{\it Acknowledgements --} I thank Prof. Vilenkin for comments (concerning Eq.~(\ref{eq:H})), and for pointing out possible issues with bounds on $\xi$, and the $\Lambda \lp^2 = 0$ case. I would also like to thank Qingdi Wang for clarifying several aspects of the analysis in \cite{unruh-wang-cc-flucs, unruh-wang-cc-flucs2}, its differences with other related works, and for pointing out that version 1 of \cite{unruh-wang-cc-flucs2} contains the discussion based on geodesic deviation equation, similar to the one presented here. Comments from Naresh Dadhich and Sumanta Chakraborty (on an earlier version of this draft) and T. Padmanabhan are gratefully acknowledged.
\\
\appendix

\section{Derivation of Eq. (\ref{eq:xi-evolution})} \label{app:dev-eq}
The aim is to start with the equation
	\begin{eqnarray}
	\bm \nabla_{\bm u} \bm \nabla_{\bm u} \xi^i = R^i_{\phantom{i}abm} u^a u^b \xi^m 
	\end{eqnarray}
and derive an equation governing the magnitude of $\bm \xi$. Substituting $\bm \xi = \xi \bm n$ in the above equation gives,
	\begin{eqnarray}
	n^i {\bm \nabla^2_{\bm u} \xi} + 2 ({\bm \nabla_{\bm u} \xi}) ({\bm \nabla_{\bm u} n^i}) + \xi \bm \nabla_{\bm u} \bm \nabla_{\bm u} n^i = 
	\xi R^i_{\phantom{i}abm} u^a u^b n^m
	\nn \\
	\end{eqnarray}
Taking dot product of the above equation with $n^i$ and using $\bm n \cdot \bm n=1$ the immediately yields Eq. (\ref{eq:xi-evolution}).

\section{Derivation of Eq. (\ref{eq:xi-schw})} \label{app:schw}
The congruence here is that of radial geodesics that start from rest at $r=R$. Their 4-velocities, in standard Schwarzschild coordinates, are then given by 
	\begin{equation}
	\bm u = \frac{\sqrt{f(R)}}{f(r)} {\bm \partial_t} - \sqrt{f(R) - f(r) } {\bm \partial_r}
	\end{equation}
	where $f(r)=1-a/r$ and $a=2GM/c^2$.
	Since we are effectively working in the two dimensional sector and ignoring $\theta, \phi$ coordinates, the unit vector $\bm n$ can be determined from the conditions 
	$\bm u \cdot \bm n=0$ and $\bm n \cdot \bm n=1$; it turns out to be
	\begin{equation}
	\bm n = \frac{\sqrt{f(R) - f(r)}}{f(r)} {\bm \partial_t} - \sqrt{f(R)} {\bm \partial_r}
	\end{equation}
	Explicit computation using above expressions gives, in this case, $\nabla_{\bm u} \bm n=0$. We can now invoke Eq. (\ref{eq:xi-evolution})
		\begin{eqnarray}
		\frac{\bm \nabla^2_{\bm u} \xi}{\xi} = - R_{abcd} u^a n^b u^c n^d + \l( \bm \nabla_{\bm u} \bm n \r)^2 
		\end{eqnarray}
		with $\nabla_{\bm u} \bm n=0$ and $R_{abcd} u^a n^b u^c n^d = - a/r^3$ for Schwarzschild geometry to rewrite the above equation as
		\begin{eqnarray}
		\frac{\DM^2 \xi}{\DM \tau^2} = \frac{a}{r(\tau)^3}\; \xi
		\end{eqnarray}
		with $\tau$ being the proper time along $u^i$. We may now rewrite the above equation using (from the knowledge of the trajectory) $\eta = 2 \arccos \sqrt{r/R}$ and converting proper time derivatives to $\eta$ derivatives  using $\tau/a = (1/2)(R/a)^{3/2} (\eta + \sin \eta)$. The final result, after some simplifications, is 		Eq.~(\ref{eq:xi-schw}).

\section{Comment on ``\textit{Spacetime geometry fluctuations and geodesic deviation}" by Vieira, Ford, and Bezerra \cite{ford-etal1}} \label{app:comment}

Our analysis of rms fluctuations is along the lines of \cite{ford-etal1} (hereafter VFB). However, the relevant equations and hence the final results these authors obtain are different from those derived above. It is therefore worth clarifying the differences.
	
	 1. VFB define (using our notation) a separation velocity $v_{\rm r} = \bm n \bm \cdot \nabla_{\bm u} \bm \xi \equiv \bm n \bm \cdot \nabla_{\bm \xi} \bm u$ 
	 by integrating the equation of geodesic deviation (GDE) under the assumption that separation between geodesics do not change much over the time scales of 
	 interest. Compare $v_{\rm r}$ with the numerator in Eq.~(\ref{eq:bgv-defn}); the two definitions are clearly different. Further, the VFB definition gives 
	 $v_{\rm r} = 0$ in flat spacetime, whereas in our case it will be governed by the initial expansion $\mathscr{H}(0)$, which need not be zero. 
	 Similar comments apply to VFB's result for $\xi$, which they obtain by a further integration of $v_{\rm r}$ under the same assumption of small separation.
	
	2. One may also seek a relation between $v_{\rm r}$ and $\dot \xi$, motivated by our result that the BGV definition of $\mathscr{H}$ yields $\dot \xi/
	\xi$. Again, it is not difficult to see that the two quantities have no direct relation. 
	Our evolution equation Eq.~(\ref{eq:xi-evolution}) for $\xi$ (which follows directly from the GDE) differs from the one used by VFB due to 
	the presence of the extra term $(\nabla_{\bm u} \bm n)^2$; even if this term is ignored, the solution for $\xi$ is not simply given by integrating the RHS of 
	Eq.~(\ref{eq:xi-evolution}) twice (note that the equation for $\xi$ is that of a time dependent harmonic oscillator). We have instead solved perturbatively the Riccati 
	type differential equation for $\mathscr{H}$, Eq.~(\ref{eq:hubble-evolution}), and then obtained $\xi$.
	
	3. Due to above reasons, when one compares fluctuations of $\xi$ (or even $\dot \xi = \mathscr{H} \xi$) with the results of VFB, they will in general be different. 
	The key difference is captured by the function $ \mathcal{F}(\tau, \tau')$ in the rms fluctuations Eqs.~(\ref{eq:rms-flucs}). Such a term does not 
	appear in VFB, due to the reasons mentioned in points 1 and 2 above. It is obvious that presence of this function will modify the time 
	dependence of the rms fluctuations. For fluctuations around flat spacetime (which is also the case 
	considered in VFB), $\Lambda=0$, and it is easily shown that 
	$\lim \limits_{\Lambda \to 0} \mathcal{F}(\tau, \tau') = {\l(1 + \mathscr{H}(0) \tau \r)^2}/{\l(1 + \mathscr{H}(0) \tau' \r)^2}$. Remarkably, however, the limiting behaviour 
	of rms fluctuations in this case, given by Eqs.~(\ref{eq:rms-flucs-short-time}) and (\ref{eq:rms-flucs-long-time}) above (for $\Lambda=0$), turns out to be qualitatively 	similar to those obtained by VFB for fluctuations sourced by thermal gravitons. Since we did not assume any specific source or form of the curvature fluctuations 
	to arrive at these scalings, it will be interesting to probe the reason for this. 



\end{document}